\documentclass[twocolumn,showpacs,preprintnumbers,amsmath,amssymb]{revtex4}

\usepackage{graphicx,psfrag}
\usepackage{amssymb}
\usepackage{amsmath}
\usepackage[squaren,Gray,textstyle]{SIunits}
\usepackage{dcolumn}
\usepackage{bm}

\usepackage[dvips]{color}

\def\nc{\newcommand}

\def\lsim{\mathrel{\raise.3ex\hbox{$<$\kern-.75em\lower1ex\hbox{$\sim$}}}}
\def\gsim{\mathrel{\raise.3ex\hbox{$>$\kern-.75em\lower1ex\hbox{$\sim$}}}}
\nc{\half}{\frac{1}{2}}
\nc{\shalf}{\ensuremath{\textstyle \frac{1}{2}}}

\nc{\deldag}{\mathbin{\partial\mkern-10.5mu\big/}}
\nc{\kdag}{\mathbin{k\mkern-10mu\big/}}
\nc{\Pdag}{\mathbin{P\mkern-10mu\big/}}

\begin{document}


\title{On the stability of spherically symmetric spacetimes in metric $f(R)$ gravity}
\author{Kimmo Kainulainen}
   \email{Kimmo.Kainulainen@phys.jyu.fi}
\author{Daniel Sunhede}
   \email{Daniel.Sunhede@phys.jyu.fi}
   
\affiliation{Dept.~of Physics, P.O.~Box 35 (YFL),
   		FI-40014 University of Jyv\"askyl\"a, Finland and\\
		Helsinki Institute of Physics and Dept.~of Physical Sciences,
		P.O.~Box 64, FI-00014 University of Helsinki, Finland;
		}

\date{\today}

\begin{abstract}
We consider stability properties of spherically symmetric spacetimes of stars in metric $f(R)$ gravity. We stress that these not only depend on the particular model, but also on the specific physical configuration. Typically configurations giving the desired $\gamma_{\rm PPN} \approx 1$ are strongly constrained, while those corresponding to $\gamma_{\rm PPN} \approx 1/2$ are less affected. Furthermore, even when the former are found strictly stable in time, the domain of acceptable static spherical solutions typically shrinks to a point in the phase space. Unless a physical reason to prefer such a particular configuration can be found, this poses a naturalness problem for the currently known metric $f(R)$ models for accelerating expansion of the Universe.
\end{abstract}


\maketitle

%
%

\section{Introduction}
\label{sec:Intro}

The observation that the expansion of the Universe appears to be accelerating~\cite{astier,spergel} has provoked discussion of a number of models for extended gravity involving nonlinear interactions in the Ricci scalar $R$:
\begin{equation}
	S = \frac{1}{2\kappa} \int {\rm d}^{4}x \sqrt{-g} [R + f(R)]
		+ S_{\rm m} \,.
\label{eq:action}
\end{equation}
Here $\kappa \equiv 8\pi G$, $S_{\rm m}$ is the usual matter action and $f(R)$ describes the new physics in the gravity sector; setting $f(R) = -2\Lambda$ corresponds to the canonical Einstein-Hilbert action in General Relativity (GR) with a cosmological constant $\Lambda$. The idea is that if cosmological data could be fitted by the use of some nontrivial function $f(R)$, one might avoid the theoretical difficulties and fine-tuning issues related to a pure cosmological constant. However, it has been shown that when understood as a {\em metric} theory, the action (\ref{eq:action}) can lead to predictions that are not consistent with Solar System measurements~\cite{chiba,erickcek,Kainulainen:2007bt}. While observations require a parameter $|\gamma_{\rm PPN}-1|\lesssim 10^{-4}$~\cite{obsongamma} in the Parametrized Post-Newtonian (PPN) formalism, the value predicted in metric $f(R)$ theories is typically $\gamma_{\rm PPN} \approx 1/2$. This is certainly the case~\cite{chiba,erickcek,Kainulainen:2007bt} for the first simple $f(R)$ models suggested in the literature~\cite{vollick,carroll}. It is however difficult to make a completely generic prediction of this result and there have been many arguments both for~\cite{chiba,erickcek,Kainulainen:2007bt,metricFail} and against~\cite{metricPass,Zhang:2007ne,Hu:2007nk,Nojiri:2007as, Nojiri:2007cq, Clifton:2008jq} metric $f(R)$ gravity failing Solar System tests. In particular, more complicated $f(R)$ functions have since been suggested which claim to yield $\gamma_{\rm PPN} \approx 1$~\cite{Zhang:2007ne,Hu:2007nk,Nojiri:2007as, Nojiri:2007cq}.

In this paper we set up the conditions which the function $f(R)$ must fulfill, so that a solution to the field equations which is compatible with Solar System observations exists, in particular with $\gamma_{\rm PPN} \approx 1$. However, we will also argue that the mere existence of such a solution does not imply that a model is consistent with observations. Since metric $f(R)$ gravity is a fourth-order theory, spacetime geometry and matter are not in as strict a correspondence as in General Relativity; depending on the boundary conditions on the metric, a given matter distribution can be consistent with different static spacetimes and with different values of $\gamma_{\rm PPN}$. Moreover, no physical principle tells us that only the boundary conditions corresponding to $\gamma_{\rm PPN} \approx 1$ solutions should be acceptable. The question is then, which solutions are the most natural ones? How plausible is it that the collapse of a protostellar dust cloud leads to the formation of the spacetime observed in the Solar System?  To answer these questions one would ideally like to study the full dynamical collapse, and given a domain of reasonable initial conditions, determine the attractor in the configuration space of possible solutions. This computation is beyond the scope of this paper however. We will instead approach the problem by studying how the time stability argument constrains the phase space of configurations with the desired properties.

The conditions that a generic metric $f(R)$ model should satisfy in order to yield acceptable solutions are: first, the Ricci scalar should closely follow the trace of the energy-momentum tensor inside a changing matter distribution, where at the same time the dimensionless quantities $f/R$ and $F \equiv \partial f/\partial R$ should remain much smaller than 1 at regions of high density. Second, the effective mass term $m^2_R$ for a perturbation in the Ricci scalar should be positive in order to assure that the GR-like $\gamma_{\rm PPN} \approx 1$ configurations are stable in time. Third, the mass $m_R^2$ should remain small so that a finite domain of static, GR-like configurations exist. This is guaranteed if $m_R^2 \lesssim 1/r_{\odot}^2$, where $r_{\odot}$ is the radius of the Sun. If this last condition is not fulfilled, the domain of GR-like configurations shrinks to essentially a point in the phase space, while a continuum of equally good, but observationally excluded, solutions still exists. In such a case the credibility of the theory requires an argument as to why the particular GR-like configuration should be preferred. None of the models so far proposed in the literature, including Refs.~\cite{Hu:2007nk,Nojiri:2007as,Nojiri:2007cq}, satisfy all of these constraints, and we also failed to construct a model that would. Largely this failure comes from the difficulty to keep both the function $F$ and $m_R^2 \sim 1/(3F_{,R})$ small simultaneously when the Ricci scalar follows the matter distribution, $R \approx \kappa\rho$.

It should be noted that the above considerations only apply for the {\em desired} GR-like $\gamma_{\rm PPN} \approx 1$ configurations when a model is tuned to mimic a (very small) cosmological constant. Metric models with a true cosmological constant plus some additional sufficiently small $f(R)$ correction can be perfectly fine. Thus the above arguments do not exclude generic $f(R)$ modifications, such as might arise from quantum corrections, to the Einstein-Hilbert action. One should also note that it is in general easy to construct stable attractor solutions yielding $\gamma_{\rm PPN} \approx 1/2$ in metric $f(R)$ theory. It is the precision data from the Solar System which makes these solutions unacceptable.

The paper is organized as follows. We start by reviewing the Solar System constraints in Sec.~\ref{sec:ppn}. We consider the Dolgov-Kawasaki time instability~\cite{Dolgov:2003px} in Sec.~\ref{sec:timestab} and discuss the corresponding stability criterion for spherically symmetric configurations. Section \ref{sec:paltrack} considers static configurations and the possibility for metric $f(R)$ gravity to follow stable, GR-like solutions that are compatible with Solar System constraints. We find that the condition for finding a finite \emph{domain} of boundary conditions giving rise to a GR-like metric is nearly orthogonal to the time stability condition. Finally, Sec.~\ref{sec:summary} contains our conclusions and discussion.

%
%

\section{Solar System constraints and the Palatini track}
\label{sec:ppn}

Let us begin by reviewing the main constraints from the Solar System observations on static solutions in metric $f(R)$ gravity. Varying the action (\ref{eq:action}) with respect to the metric gives the equation of motion:
\begin{eqnarray}
	(1+F) R_{\mu \nu} - \frac{1}{2} (R+f) g_{\mu \nu} && \nonumber \\ 
    - \nabla_\mu\nabla_\nu F+ g_{\mu \nu}\Box F & = & \kappa T_{\mu \nu} \,,
\label{eq:eom}
\end{eqnarray}
where $F \equiv f_{,R} = \partial f/\partial R$ and $\Box = g^{\mu\nu}\nabla_\mu\nabla_\nu$.  Taking the trace of this equation one finds:
\begin{equation}
	\Box F - \frac{1}{3}(R(1-F) + 2f) = \frac{1}{3}\kappa T \,.
\label{eq:trace}
\end{equation}
If $F\rightarrow 0$ and $f\rightarrow R$ this equation reduces to the standard algebraic GR relation between the Ricci scalar and the trace of the energy-momentum tensor $T$.  In a generic metric $f(R)$ theory $R$ is a dynamical variable however, and the theory may exhibit an instability which we will discuss in the next section. Assuming a static, spherically symmetric metric $g_{\mu \nu}$,
\begin{equation}
	ds^2 \equiv g_{\mu \nu} x^{\mu} x^{\nu} =
		-e^{A(r)}{\rm d}t^2 + e^{B(r)}{\rm d}r^2 + r^2{\rm d}\Omega^2 \,,
\label{eq:metric}
\end{equation}
the full field equations (\ref{eq:eom}) reduce to the following source equations for the metric functions $A$ and $B$ in the weak field limit (to first order in small quantities):
\begin{eqnarray}
	(rB)' & \approx & \kappa \rho r^2  \bigg( 1
			- \frac{1}{3}\bigg[ \frac{1+3F}{1+F}
				- \frac{R}{\kappa\rho}\frac{1 + \frac{F}{2}+\frac{f}{2R}}{1+F}
			\bigg] \bigg) \nonumber \\
		& & - \gamma rA' \,,
\label{eq:sourceB} \\
	A' & \approx & \frac{1}{1+\gamma} \left( \frac{B}{r}
		- \frac{r}{2(1+F)}\bigg[ FR - f + \frac{4}{r}F' \bigg] \right) \,,\qquad
\label{eq:sourceA}
\end{eqnarray}
where $\gamma \equiv rF'/2(1+F)$, a prime refers to a derivative with respect to $r$, and we have neglected pressure so that $T \approx -\rho$. The parameter $\gamma$  and the terms in the square brackets highlight the deviation from General Relativity.  The value of $\gamma_{\rm PPN} \approx - B/A$ far away from a gravitational source depends on the continuous evolution of $A$ and $B$ throughout the Sun. It is particularly sensitive to the evolution through the core where the density is the highest. Hence, to obtain $\gamma_{\rm PPN} \approx 1$ and the correct gravitational strength in the Solar System, the only solution, not obviously dependent on an enormous amount of fine-tuning~\footnote{By this we mean that it is in principle possible that an evolution of $B/A$ quite different from the one in GR, could actually lead to the same value for the exterior of a star. However, away from the GR track the outcome becomes sensitive to the form of the density profile, which leads to even more uncertainties as to how the actual dynamical gravitational collapse would proceed.}, is that the extra terms in Eqns.~(\ref{eq:sourceB}-\ref{eq:sourceA}) must remain small throughout the interior of the Sun. Now, if the extra terms can be neglected in the $B'$ equation, one finds that $B \lesssim 10^{-6}$ throughout the interior of the Sun~\cite{Kainulainen:2007bt}. It then becomes clear from the $A'$ equation that $f/R$, $F$, and $rF'$ need to be very small compared to 1. However, to make the correction vanish in the $B'$ equation one in addition needs to require that the Ricci scalar traces the matter density $R/\kappa \rho \approx 1$. So, barring perhaps some fantastic fine-tunings, the only possibility is that one finds a configuration for which: 
\begin{equation}
	F \ll 1 \,, \quad
	f/R \ll 1 \,, \quad {\rm and} \quad
	R \approx \kappa \rho \,.
\label{eq:conditions}
\end{equation}
Note that this limit was also discussed in Ref.~\cite{Hu:2007nk}. Here we see that the above conditions are a necessary requirement for fulfilling the local gravity constraints.

In GR the equation $R = -\kappa T$ is of course exact (remember that we are neglecting pressure throughout so that $T\approx -\rho$), but this is in general very difficult to arrange in a metric $f(R)$ model. The problem lies in the dynamical nature of the Ricci scalar in metric $f(R)$ gravity. To see this, consider the static trace equation (\ref{eq:trace}) in the weak field limit:
\begin{eqnarray}
	F'' + \frac{2}{r}F' & = & \frac{1}{3} \big(R - \kappa \rho - FR + 2f  \big)
		\nonumber \\
		& \equiv & \frac{1}{3} \big(\Sigma(F) - \kappa \rho \big) \,,
\label{eq:traceWeak}
\end{eqnarray}
where we have again assumed that pressure is negligible. Assume now that $F=F_0$ at the center of the Sun. If the nonlinear term $\Sigma(F)$ is small compared to $\kappa \rho$, then the solution for $F$ 
becomes:
\begin{equation}
	F(r) = - \int_0^r {\rm d}r' \frac{2G m(r')}{3r'^2} + F_0 \,,
\label{eq:Fsol}
\end{equation}
where $m(r) \equiv \int_0^r {\rm d}r' 4\pi r'^2 \rho$. In the case of the Sun this implies that $F$ evolves only little: $|F(r)-F_0| \lesssim 10^{-6}$~\cite{Kainulainen:2007bt}. The solution (\ref{eq:Fsol}) is in general not compatible with $R \approx \kappa \rho$ as required by the conditions (\ref{eq:conditions}).

Let us now set $\Sigma(F_0) = \kappa \rho_0$ at $r = 0$. If the gradients somehow remain small throughout the evolution, then the solution follows the Palatini trace equation:
\begin{equation}
	\Sigma(F) = R(1-F) + 2f = 8\pi G \rho \,.
\label{eq:tracePal}
\end{equation}
For small $F$ and $f/R$ this evolution {\em would} be consistent with the condition $R\approx \kappa \rho$ and, since we are following the Palatini track, give  $\gamma_{\rm PPN} \approx 1$ (see {\em e.g.~}\cite{solarPal}). 
Whether such a solution actually exists is more difficult to prove. However, one can study under which conditions such a solution, if it exists, would be an attractor and whether it would also be sufficiently stable in time.

Finally, let us note that if $\Sigma(F) \ll \kappa \rho$ so that $F$ is given by Eqn.~(\ref{eq:Fsol}), small $F$ and $f/R$ result in a different class of solutions with $R/\kappa \rho \ll 1$. In this case the field Eqns.~(\ref{eq:sourceB}-\ref{eq:sourceA}) reduce to
\begin{eqnarray}
	(rB)' & \approx & \frac {2}{3}\kappa \rho r^2 \,,
\label{eq:sourceB2} \\
	A' & \approx & \frac{B}{r} - 2F' \,.
\label{eq:sourceA2}
\end{eqnarray}
It is easy to show that together with Eqn.~(\ref{eq:Fsol}) these give $\gamma_{\rm PPN} \approx 1/2$.

%
%

\section{Time stability}
\label{sec:timestab}

\begin{figure}[!t]  
    \begin{center}
    \includegraphics[width=8cm]{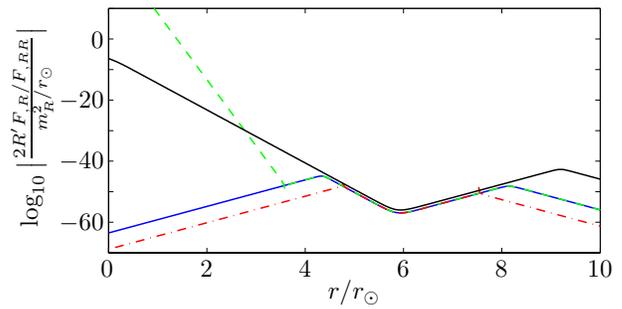}%
    \end{center}
    \caption{The gradient term proportional to $\delta R'$ in Eqn.~(\ref{eq:delR}), normalized to $m_R^2/r_{\odot}$, for various $f(R)$ models ($R = \kappa \rho$): $-\mu^4/R$ (solid blue), $-\mu^4/R + \alpha R^2/\mu^2$ (dashed green), Hu \& Sawicki (dot-dashed red)~\cite{Hu:2007nk}, and $\alpha R \log{(R/\mu^2)}$ (solid black). The actual density profile used in all figures corresponds to the known density profile of the Sun with a central density of $150$ g/cm$^3$ and with a roughly exponential dependence on $r$. We have also superimposed a constant dark matter distribution on the profile of the Sun with $\rho_{\rm DM} = 0.3$ GeV/cm$^3$.}
    \label{fig:gradient}
\end{figure}

Let us first consider the time stability of spherically symmetric configurations in generic $f(R)$ models. Perturbing around some arbitrary configuration, $R(r) \rightarrow \widetilde{R}(r,t) = R(r) + \delta R(r,t)$, and expanding to first order in $\delta R$, $\delta R'$ and $\dot{\delta R}$, where the prime refers to a derivative with respect to $r$ and the dot to a derivative with respect to $t$, one can write the trace equation (\ref{eq:trace}) in the following form in the weak field limit:
\begin{eqnarray}
	(\partial_t^2 - \vec{\nabla}^2) \delta R & = &
		- m_R^2\delta R
		+ 2\frac{F_{,RR}}{F_{,R}}R'\delta R' 
\nonumber \\		
		& & + \vec{\nabla}^2 R + \frac{1}{3F_{,R}}\Delta
		+ \frac{F_{,RR}}{F_{,R}} (R')^2 \,, \quad
\label{eq:tracePert}
\end{eqnarray}
where $\Delta \equiv -\kappa T - R(1-F) - 2f$ and
\begin{equation}
	m_R^2 \equiv \frac{1}{3F_{,R}}(1-F-\varepsilon) \,,
\label{eq:mR}
\end{equation}
with
\begin{eqnarray}
	\varepsilon \phantom{.} \delta R & \equiv &
		R(\widetilde{F}-F) - 2(\widetilde{f} - f)
			+ \bigg(1-\frac{\widetilde{F}_{,R}}{F_{,R}} \bigg)\Delta
		\nonumber \\
		&& {}+ 3 F_{,RR}\bigg(\frac{\widetilde{F}_{,RR}}{F_{,RR}}
						- \frac{\widetilde{F}_{,R}}{F_{,R}}\bigg)(R')^2 \,.
\label{eq:epsilon}
\end{eqnarray}
Here a tilde is used to denote that a quantity is perturbed, {\em i.e.~}it is a function of $\widetilde{R}(r,t)$ as opposed to the background value $R(r)$. Assuming that the configuration $R(r)$ we are perturbing around is a solution to the static equation, the second line in Eqn.~(\ref{eq:tracePert}) drops out. Moreover, for most cases the gradient term proportional to $\delta R'$ is completely negligible inside a stellar object and can be dropped as well. See Fig.~\ref{fig:gradient} for some examples. The behavior of the perturbation around a static, spherically symmetric solution is thus governed by the equation
\begin{figure}[!t]  
    \begin{center}
    \includegraphics[width=8cm]{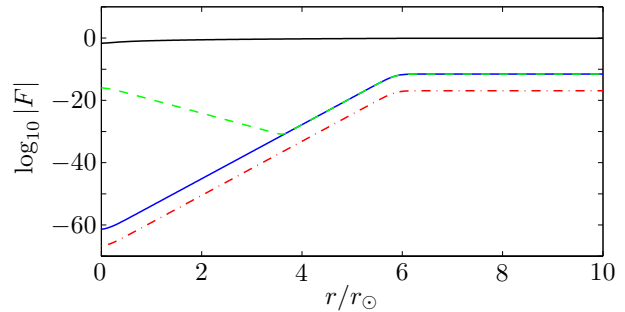}%
    \end{center}
    \caption{The parameter $F$ given as a function of the radius for various $f(R)$ models ($R = \kappa \rho$): $-\mu^4/R$ (solid blue), $-\mu^4/R + \alpha R^2/\mu^2$ (dashed green), Hu \& Sawicki (dot-dashed red)~\cite{Hu:2007nk}, and $\alpha R \log{(R/\mu^2)}$ (solid black).}
    \label{fig:F}
\end{figure}
\begin{eqnarray}
	(\partial_t^2 - \vec{\nabla}^2) \delta R & = &
		- m_R^2 \delta R \,.
\label{eq:delR}
\end{eqnarray}
Note that the mass $m_R^2$ only depends on the background value of the Ricci scalar $R(r)$.

Table \ref{table1} and \ref{table2} show the components of $m_R^2$ in some particular models and the corresponding parameter values used in all figures are displayed in Table \ref{table3}. As discussed in the previous section, $F$ needs to be small compared to one for GR-like configurations (see Fig.~\ref{fig:F}, we will discuss this constraint further below). When this is the case one typically finds that also $\varepsilon \ll 1$, so that $m_R^2 \approx 1/3F_{,R}$.  It then follows that if $F_{,R} < 0$, then $m_R^2 < 0$ and the coefficient of $\delta R$ is negative for the configuration in question, so that system exhibits an instability. This is the instability first found by Dolgov and Kawasaki in the context of an $f(R) = -\mu^4/R$ model~\cite{Dolgov:2003px} (see Ref.~\cite{Faraoni:2006sy} for a more general case). It is important to note that the instability  depends not only on the model, but also on the particular configuration. Certain configurations in a given model are more stable than others and the instability may even vanish in some cases.

The nature of the instability is most transparent in the special case with constant curvature. Then $m_R$ is a constant and one can obtain an exact solution for $\delta R(r,t)$. Expanding $\delta R$ in Fourier modes, one finds that a mode with wave vector $\vec{k}$ has the time dependence
\begin{eqnarray}
	\delta R_k(\vec{k},t) \sim e^{\pm i\sqrt{k^2 + m_R^2}t} \,,
\label{eq:delRsol}
\end{eqnarray}
so that for negative $m_R^2 \sim 1/3F_{,R}$, all modes with $k < |m_R|$ are unstable. This does not necessarily rule out a model however. If for example $|m_R| \sim H_0$, then the instability time is much longer than the lifetime of the Solar System and the model is safe. Moreover, whenever $|m_R|^{-1}$ is much larger than the size of the physical system under consideration, only modes corresponding to scales much larger than the system are unstable and this can not alter its local geometry.

Now, assume that we have a GR-like solution, such that $R \sim -\kappa T \approx 8\pi G \rho$. One then has
\begin{equation}
	\frac{R}{\mu^2} \sim 10^{29} \left(\frac{\Lambda}{\mu^2} \right)
		\left(\frac{\rho}{{\rm g}/{\rm cm}^3}\right) \,,
\label{eq:Rmagn}
\end{equation}
where we have used $\Lambda \approx 0.73 \kappa \rho_{\rm crit}$. Hence, for a pure $-\mu^4/R$ model the mass squared is on the order of
\begin{equation}
m_R^2 \sim -(10^{-26} \textrm{ s})^{-2}\left(\frac{\Lambda}{\mu^2} \right)^2
		\left(\frac{\rho}{{\rm g}/{\rm cm}^3}\right)^3 \,.
\end{equation}
This system is violently unstable at normal densities for 
\begin{widetext}
\center{
\begin{table}[!htb] 
\begin{tabular}{cc|ccc|cc}
	$\displaystyle f(R)$ & & &
	$\displaystyle F$ & & &
	$\displaystyle 1/3F_{,R}$ \\
\hline
& & & & & &\\
	$\displaystyle -\frac{\mu^4}{R}$ & & &
	$\displaystyle \frac{\mu^4}{R^2}$ & & &
	$\displaystyle -\frac{R^3}{6\mu^4}$ \\
& & & & & &\\
	$\displaystyle -\frac{\mu^4}{R} + \alpha \frac{R^2}{\mu^2}$ & & &
	$\displaystyle \frac{1}{\bar{R}^2} + 2\alpha\bar{R}$ & & &
	$\displaystyle \frac{\mu^2}{6(\alpha - 1/\bar{R}^3)}$ \\
& & & & & &\\
	$\displaystyle -\mu^2 \frac{c_1(R/\mu^2)^n}{c_2(R/\mu^2)^n + 1}$ & & &
	$\displaystyle -\frac{c_1n}{c_2^2\bar{R}^{n+1}(1+1/c_2\bar{R}^n)^2}$ & & &
	$\displaystyle
		\mu^2 \frac{c_2^2 \bar{R}^{n+2}}{3c_1n(n+1)} \phantom{.}
			\frac{(1 + 1/c_2\bar{R}^n)^3}{1-\frac{n-1}{n+1}/c_2\bar{R}^n}$ \\
& & & & & &\\
	$\displaystyle \alpha R \log{\frac{R}{\mu^2}}$ & & &
	$\displaystyle \alpha \Big(1+\log{\frac{R}{\mu^2}}\Big)$ & & &
	$\displaystyle \frac{R}{3\alpha}$ \\
\end{tabular}
\caption{The function $F$ and the dominant term of the mass squared $m_R^2 \equiv (1- F-\varepsilon)/3F_{,R}$ for different $f(R)$ models, where $\bar{R} \equiv R/\mu^2$.}
\label{table1}
\end{table}
}
\center{
\begin{table}[!htb] 
\begin{tabular}{cc|cc}
	$\displaystyle f(R)$ & & &
	$\displaystyle \varepsilon$ \\
\hline
& & & \\
	$\displaystyle -\frac{\mu^4}{R}$ & & &
	$\displaystyle -3\Big(1+\frac{\kappa T}{R}\Big) + \frac{5\mu^4}{R^2}
		+ 3 [3F_{,R}] \Big(\frac{R'}{R} \Big)^{\! 2}$ \\
& & & \\
	$\displaystyle -\frac{\mu^4}{R} + \alpha \frac{R^2}{\mu^2}$ & & &
	$\displaystyle
		-\frac{2\alpha^2\bar{R} + 2\alpha/\bar{R}^2 -
				3(1+\bar{\kappa}\bar{T}/\bar{R})/\bar{R}^3 + 5/\bar{R}^5}{
			\alpha - 1/\bar{R}^3}
		- 3 \frac{4\alpha-1/\bar{R}^3}{\bar{R}^3(\alpha-1/\bar{R}^3)^2}
		[3F_{,R}] \Big(\frac{R'}{R} \Big)^{\! 2}$ \\
& & & \\
	$\displaystyle -\mu^2 \frac{c_1(R/\mu^2)^n}{c_2(R/\mu^2)^n + 1}$ & & &
	$\displaystyle
		-\frac{n+2}{D}\left( 1 + \frac{\bar{\kappa}\bar{T}}{\bar{R}}
					 - \frac{2c_1}{c_2\bar{R}(1+1/c_2\bar{R}^n)^2}
		\right)
		- \frac{n+2}{D^2}
		[3F_{,R}] \Big(\frac{R'}{R} \Big)^{\! 2}$ \\
& & & \\
	$\displaystyle \alpha R \log{\frac{R}{\mu^2}}$ & & &
	$\displaystyle -1 - \frac{\kappa T}{R} - 3\alpha\log{\frac{R}{\mu^2}}
		+ [3F_{,R}] \Big(\frac{R'}{R} \Big)^{\! 2}$ \\
\end{tabular}
\caption{The parameter $\varepsilon$ for different $f(R)$ models, where $D \equiv (1-\frac{n-1}{n+1}/c_2\bar{R})(1+1/c_2\bar{R}^n)$ and a bar indicates that a quantity is dimensionless and measured in units of $\mu$, {\it e.g.~}$\bar{R} \equiv R/\mu^2$. In the third model above, originally suggested in Ref.~\cite{Hu:2007nk}, we have only kept the leading terms in $\varepsilon$.}
\label{table2}
\end{table}
}
\end{widetext}
all scales larger than $\sim 10^{-18}{\rm~m}$,  if $\mu$ is fixed to account for the present accelerating expansion of the Universe.

As was pointed out by Dick~\cite{Dick:2003dw} and later discussed by Nojiri \& Odintsov~\cite{Nojiri:2003ft}, adding a conformal term $\alpha R^2/\mu^2$ can stabilize this system; for $f(R) = -\mu^4/R + \alpha R^2/\mu^2$ the previous approximation for the mass reads:
\begin{equation}
	m_R^2 \sim -\frac{R^3}{6\mu^4}\left(\frac{1}{1-\alpha R^3/\mu^6}\right) 
	\sim \frac{\mu^2}{\alpha} \,,
\label{eq:mRmagn}
\end{equation}
where the last step assumes that $\alpha R^3/\mu^6 > 1$. If $\alpha \sim 1$ this may be true even for $R \sim \mu^2$ so that one always finds a very small positive mass $m_R^2 \sim \mu^2/\alpha \sim (10^{18}\textrm{ s})^{-2}$.

However, the above stabilization mechanism runs into problems with the conditions in Eqn.~(\ref{eq:conditions}). Indeed, for $\alpha \sim 1$ and $R \approx \kappa \rho \gg \mu^2$ one has
\begin{equation}
F = \frac{\mu^4}{R^2} + \alpha \frac{R}{\mu^2} \gg 1\,,
\end{equation}
so that the configuration would clearly not be GR-like~\footnote{In general, $F>1$ implies that the leading approximation in Eqn.~(\ref{eq:mRmagn}) is no longer valid. However, it still holds for $f(R) = -\mu^4/R + \alpha R^2/\mu^2$, since the contribution of the conformal term in $F$ will cancel with the leading term in $\varepsilon$ where additional terms remain small compared to one.}.
The problem is that changing $F$ modifies the effective strength of the gravitational constant $G_{\rm eff} = G/(1+F)$, which controls the buildup of the gravitational potential inside the star. In fact, for $\alpha \sim 1$ the effect is so strong that it would weaken the gravitational force so much as to prohibit the growth of any density contrasts much above the critical density. This argument can be turned around to a constraint: in order for the function $F$ to remain small inside the densest objects we have reasonably accurate information on, the neutron stars, one has to have $\alpha \kappa \rho_{\rm nucl}/\mu^2 \ll 1$, where $\rho_{\rm nucl}$ is the nuclear density~\footnote{The loophole to this argument is if one only follows $R \approx \kappa \rho$ inside the Sun, but that the Ricci scalar is allowed to significantly deviate from this relation inside neutron stars. In such a case one could maintain $F \ll 1$ for much larger values of $\alpha$, resulting in a smaller $m_R^2$ for the Sun. This scenario seems very contrived however.}.
Since $\mu^2 \sim \kappa \rho_{\rm crit}$ we find that 
\begin{equation}
\alpha \lesssim \frac{\rho_{\rm crit}}{\rho_{\rm nucl}} \sim 10^{-45} \,.
\end{equation}
This is quite a stringent constraint, but it does not rule out the model based on the required time stability. Indeed, for example with $\alpha = 10^{-47}$ one has $\alpha (\kappa \rho )^3/\mu^6 = 1$ when $\rho \sim 10^{16}\rho_{\rm crit}$. For any density higher than this value, the system is stable in time with a very {\em large} positive mass squared given by the formula in Eqn.~(\ref{eq:mRmagn}): $m_R^2 \sim \mu^2/\alpha \sim (10^{-6} {\rm~s})^{-2}$. There is a caveat to this argument however, since for these parameters the gradient term proportional to $\delta R'$ becomes very large inside the Sun (see Fig.~(\ref{fig:gradient})) and the simplified equation (\ref{eq:delR}) can no longer be trusted.

\begin{figure}[!t]  
    \begin{center}
    \includegraphics[width=8cm]{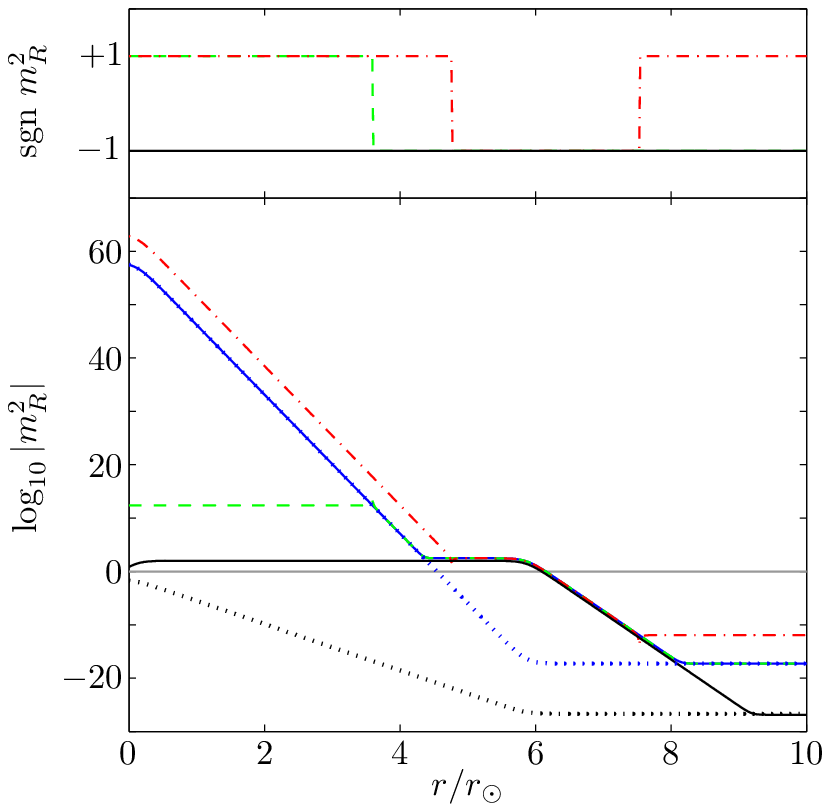}%
    \end{center}
    \caption{The mass squared $m_R^2$ in units $1/r_{\odot}^2$ for various $f(R)$ models ($R = \kappa \rho$): $-\mu^4/R$ (solid blue), $-\mu^4/R + \alpha R^2/\mu^2$ (dashed green), Hu \& Sawicki (dot-dashed red)~\cite{Hu:2007nk}, and $\alpha R \log{(R/\mu^2)}$ (solid black). The horizontal solid gray line corresponds to the limit $1/r_{\odot}^2$. For the $-\mu^4/R$ and $\alpha R\log{(R/\mu^2)}$ models we have also plotted $1/3F_{,R}$ with dotted blue and dotted black lines, respectively. The upper panel displays the corresponding sign of the mass squared where we have excluded the $-\mu^4/R$ model for which $m_R^2$ is strictly negative.}
    \label{fig:mR2}
\end{figure}

The complete mass squared function for the model with a fine-tuned conformal $\alpha R^2$ term, using the exact expression (\ref{eq:mR}), is shown in Fig.~\ref{fig:mR2} (dashed green curve). The lower panel displays the absolute value $|m_R^2|$ and the upper panel the sign of $m_R^2$. In the model at hand the mass would remain large and \emph{positive} throughout the entire interior of the Sun, which is the necessary condition for time stability. The GR-like configuration does become unstable at low densities, but this would not necessarily change the value of $\gamma_{\rm PPN}$ in the Solar System.

\begin{table}[!htb] 
\begin{tabular}{cc|cc}
	$\displaystyle f(R)$ & & &
	Parameter values \\
\hline
& & & \\
	$\displaystyle -\frac{\mu^4}{R}$ & & &
	$\displaystyle \mu^2 = 4\Lambda/\sqrt{3}$ \\
& & & \\
	$\displaystyle -\frac{\mu^4}{R} + \alpha \frac{R^2}{\mu^2}$ & & &
	$\displaystyle \mu^2 = 4\Lambda/\sqrt{3} \,, \quad \alpha = 10^{-47}$ \\
& & & \\
	$\displaystyle -\mu^2 \frac{c_1(R/\mu^2)^n}{c_2(R/\mu^2)^n + 1}$ & & &
	$\displaystyle \mu^2 = (8315 {\rm~Mpc})^{-1} \,, \quad n = 1\,,$ \\
& & & $\displaystyle c_1/c_2 = 6 \times 0.76/0.24 \,,$ \\
& & & $\displaystyle \phantom{\Bigg|}c_1/c_2^2 = 10^{-6}\times41^{n+1}/n $ \\
& & & \\
	$\displaystyle \alpha R \log{\frac{R}{\mu^2}}$ & & &
	$\displaystyle \mu^2 = 4\Lambda e^{(1-\alpha)/\alpha} \,, \quad \alpha = 1/\log{10^{32}}$ \\
\end{tabular}
\caption{Chosen parameter values for the different $f(R)$ models in Figs.~\ref{fig:gradient}-\ref{fig:mR2}, where $\Lambda = 0.73 \kappa \rho_{\rm crit}$. For the third model, originally suggested in Ref.~\cite{Hu:2007nk}, we have used values given in the original publication. Note that a value $n=4$, which was also discussed in~\cite{Hu:2007nk}, would result in an even larger value of $m_R^2$ in this scenario.}
\label{table3}
\end{table}
Fig.~\ref{fig:mR2} also displays $m_R^2$ for several other models (for parameter values used in each model see Table~\ref{table3}): the solid blue line represents the simple $-\mu^4/R$ model, which has a very large negative mass inside the Sun, and the dash-dotted red curve shows the mass function in a model by Hu \& Sawicki (HS)~\cite{Hu:2007nk}. The HS model fulfills the conditions (\ref{eq:conditions}) by construction, and its very large positive mass guarantees time stability.  The fact that $m_R^2$ becomes negative around $r \sim 6r_{\odot}$ in the HS model is caused by the $\varepsilon$ term in the complete expression (\ref{eq:mR}), but this does not necessarily have any effect on $\gamma_{\rm PPN}$. Moreover, this feature is sensitive to the particular form of the exterior density profile (where we have neglected for example the Solar wind) and it is not important for our main results. Overall one sees that the expression $m_R^2 \approx 1/3F_{,R}$ is a very good approximation for the first three models described in Table~\ref{table1}, except for a small region around $r \sim 6r_{\odot}$ where the $\varepsilon$ term may come into play. For the logarithmic model this approximation is only good for very low densities and we will discuss this in more detail in section \ref{sec:paltrack}.

We can summarize this section as follows: the stability of a static, spherically symmetric GR-like configuration with $R \approx \kappa \rho$ is predominantly governed by the mass term $m_R^2 \sim 1/3F_{,R}$.  If $F_{,R} < 0$, all perturbations with wavelengths larger than $1/|m_R|$ will be unstable so that for a large mass $|m_R|$, the curvature inside a stellar object will evolve rapidly before some nonperturbative effect stabilizes the system. Hence, in order for a model to be stable, $m_R^2$ should be positive throughout the Sun for GR-like configurations~\footnote{Configurations with a small enough negative $m_R^2$ can be accepted as well. The minimum requirement is then that at least all perturbations with wavelengths smaller than the size of the physical system under consideration should remain stable.}.
Both the model with a fine-tuned conformal $\alpha R^2$ term (apart from the caveat mentioned above) and the HS scenario do satisfy all constraints discussed so far.

%
%

\section{An upper bound on $m_R^2\,$?}
\label{sec:paltrack}

As mentioned in the introduction, a given matter distribution can be consistent with many different static geometries, depending on how the boundary conditions are defined at the center of the star. One always requires that the exterior metric is asymptotically flat and so different solutions are characterized by different values of $\gamma_{\rm PPN}$. There appears to be no {\em a priori} preference of one solution to another and indeed the question is: what is the most probable configuration to arise through gravitational collapse? Lacking a dynamical calculation we are restricted here to study how special the GR-like solutions are in the phase space.

Consider a static solution $R(r) = R_T(r) + \delta(r)$ where $R_T$ is the solution to the Palatini trace equation and $\delta/R_T \ll 1$ so that $R$ remains very close to the Palatini track. Note that the function $\delta(r)$ is not a true perturbation since $R_T(r)$ is not a solution to the complete metric trace equation (\ref{eq:trace}). However, one can easily obtain the equation governing $\delta$ via Eqn.~(\ref{eq:tracePert}), giving
\begin{eqnarray}
	\vec{\nabla}^2 \delta & = &
		\frac{1}{3F_{,R}}(1-F-\varepsilon)\delta
		- 2\frac{F_{,RR}}{F_{,R}}R'_T\delta' \nonumber \\
	& & - \vec{\nabla}^2 R_T - \frac{F_{,RR}}{F_{,R}} (R'_T)^2 \,,
\label{eq:del}
\end{eqnarray}
where $F$ and its derivatives are functions of the ``background'' value $R_T(r)$. Similarly, the ``perturbed'' quantities in the definition for $\varepsilon$ are functions of $R(r) = R_T(r) + \delta(r)$.

In analogy with the above analysis for $\delta R(r,t)$, let us consider a constant density object so that $R_T = {\rm const.}$, giving
\begin{eqnarray}
	\vec{\nabla}^2 \delta & = & m_R^2 \delta \,.
\label{eq:delConst}
\end{eqnarray}
The mass term $m_R^2$ is of course the same mass that appears in the equation for $\delta R(r,t)$. Now, for $m_R^2 < 0$, the solution for $\delta(r)$ is decaying so that the Palatini track acts as an attractor for the solution $R(r)$. This is exactly the behavior that was demonstrated by solving the full field equations in Ref.~\cite{Kainulainen:2007bt} for a $f(R) = -\mu^4/R$ model. However, if $m_R^2 > 0$, the solution for $\delta(r)$ will also contain a growing component:
\begin{equation}
	\delta(r) = \frac{C_1}{r}e^{+m_R r} + \frac{C_2}{r}e^{-m_R r}\,.
\label{eq:delSol}
\end{equation}
The fine-tuning problem we have to face is manifest from this equation: the time stability argument of the preceding section requires that $m_R^2>0$, so that Eqn.~(\ref{eq:delSol}) with its growing mode is what describes the deviation away from GR-like solutions.

From Sec.~\ref{sec:timestab} we already know that setting up a GR-like configuration requires fine-tuning at the center of the star. However, what Eqn.~(\ref{eq:delSol}) implies is worse: starting from $r=0$ we could always choose a boundary condition $(F_0,F'_0)$ that kills the growing mode, but as any perturbation around this solution would be exponentially enhanced, the boundary condition must be set with an incredible precision when $m_R^2$ is large. Numerically such a solution can always be found by use of a differentiation method that kills the growing mode as was done in Ref.~\cite{Hu:2007nk}. However, different choices of boundary conditions would lead to other physically, but not observationally, acceptable spacetimes. For example, if one starts from a point a little off from the GR track, $\delta$ initially grows exponentially pulling the solution away from $R \approx \kappa \rho$. Then the nonlinear terms typically become negligible in Eqn.~(\ref{eq:traceWeak}) and $R(F)$ starts to approach the solution of Eqn.~(\ref{eq:Fsol}). For $R/\kappa\rho \ll 1$ this limit corresponds to the evolution of $A$ and $B$ given by Eqns.~(\ref{eq:sourceB2}-\ref{eq:sourceA2}), which leads to $\gamma_{\rm PPN} \approx 1/2$. Thus, for a large $m_R^2$ the nearly singular static GR-like solution is surrounded by a continuum of equally acceptable configurations, however with observationally excluded values for $\gamma_{\rm PPN}$. 

Hence, given no physical reason to prefer a given set of boundary conditions, it would appear more natural to expect that the metric around a generic star would correspond to $\gamma_{\rm PPN}$ different from one. As stated above, to make a definitive statement would require solving the dynamical problem of collapse, but this is beyond the scope of the present work. Nevertheless we believe that we have identified a potential problem for metric $f(R)$ gravity models: for an $f(R)$ model to be credible, it is not sufficient to provide a mere proof of existence of a GR-like solution, but one should also give an argument as to why this particular solution is preferred.

The situation would be ameliorated if the growing mode is not excluded, but the length scale dictating the growth of the perturbations, $1/m_R$, is small enough. Roughly one should have
\begin{equation}
	m_R^2 \lesssim \frac{1}{r_{\odot}^2} \,,
\label{eq:constraint}
\end{equation}
throughout the Sun. However, both the HS scenario and the fine-tunded $f(R) = -\mu^4/R + \alpha R^2/\mu^2$ model fail this constraint by a large margin, as can be seen from Fig.~\ref{fig:mR2}. This is also the case for the model in Ref.~\cite{Zhang:2007ne} where a stabilizing conformal term creates a behavior very similar to the $\alpha R^2$ model discussed here. The same argument also applies to more recent models introduced in Refs.~\cite{Nojiri:2007as, Nojiri:2007cq}. These scenarios behave very similar to the HS model at late times, but were designed to also account for inflation at very high energies. For example, for the model suggested in Ref.~\cite{Nojiri:2007cq},
\begin{equation}
	f(R) = \frac{\alpha R^{m+l} - \beta R^n}{1 + \gamma R^l} \,,
\label{eq:NojOdmodel}
\end{equation}
where the authors chose $m = l = n$ for simplicity, and $n \ge 2$, one can show that the mass squared is given by~\cite{Nojiri:2007cq}
\begin{equation}
	m_{R}^2 \sim + \frac{R_I}{3n(n-1)}\left( \frac{R_I}{R} \right)^{n-1} \,.
\label{eq:NojOdmass}
\end{equation}
Here $R_I \sim (10^{15} {\rm~GeV})^2$ is set to the scale of inflation, and so this mass is enormous in comparison with the bound (\ref{eq:constraint}) inside the Sun.

The generic problem is that a small value of $m_R^2 \sim 1/3F_{,R}$ requires a large value for $F_{,R}$. However, at the same time one also needs $F \ll 1$ in order to obtain a reasonable gravitational potential. This tension is what makes it difficult to find a suitable function $f(R)$. Let us illustrate the problem further by trying to construct an explicit model by Þrst making sure that the toughest requirement is satisfied. At the center of Sun $R \approx \kappa\rho \sim 10^{31} \Lambda \sim 10^{-3}/r_{\odot}^2$, which  is much smaller than the upper limit on $m_R^2$. Thus we can take
\begin{equation}
	m_R^2 \sim +R \,.
\end{equation}
Using this together with the formula $F_{,R} \sim 1/m_R^2$, we can construct a candidate model:
\begin{equation}
	f(R) = \alpha R\log{\frac{R}{\mu^2}} \,,
\label{eq:logmodel}
\end{equation}
where $\mu^2 = 4\Lambda e^{(1-\alpha)/\alpha}$ in order to obtain the desired accelerating expansion of the Universe at present. Furthermore, demanding that $F \ll 1$ in the interior of the Sun yields $\alpha \lesssim 0.01$ so that $\mu^2 \gtrsim e^{100} \Lambda$~\footnote{However, note that since $F = \alpha(\log{(R/\mu^2)} + 1) = \alpha\log{(R/4\Lambda) - 1 + 2\alpha}$, there is some tension between getting $F \ll 1$ inside \emph{both} the Sun and a neutron star. Although this may indeed prove relevant in further considerations, it is not of importance for the discussion at hand.}.
So, curiously enough the Solar System constraints would force this model to create the desired accelerating expansion without an extremely small energy scale $\sim \sqrt{\Lambda}$. 

Unfortunately, there is a flaw in the above argumentation, since we implicitly assumed that $\varepsilon \ll 1$ by assuming $m_R^2 \sim 1/3F_{,R}$. 
This assumption was fine for the discussion in the previous sections, but it fails here. Indeed, when $F_{,R}$ is large, the gradient term proportional to $[3F_{,R}](R'/R)^2$ in $\varepsilon$ may also become large (see Table~\ref{table2}). We can estimate the size of this term using an exponential density profile for the Sun $\rho \sim \rho_0/(1 + e^{\xi(r-r_{\odot})})$, where $\xi \sim 10 r_{\odot}^{-1}$:
\begin{equation}
	\Big(\frac{R'}{R} \Big)^{\! 2} \approx \Big(\frac{\rho'}{\rho} \Big)^{\! 2}
	\sim \xi^2 \sim 100 \frac{1}{r_{\odot}^2} \,.
\end{equation}
Now, since $3F_{,R} \gtrsim r_{\odot}^2$ one gets $[3F_{,R}](R'/R)^2 \gg 1$ and our simple estimate for the mass fails. A more careful estimate in the model (\ref{eq:logmodel}) finds that the mass is dominated by the gradient term and one has
\begin{equation}
	m_R^2 \approx m_R^2\big|_{\rm grad} = -\Big(\frac{R'}{R} \Big)^{\! 2}
	\sim -100 \frac{1}{r_{\odot}^2} \,.
	\label{eq:m2RlogR}
\end{equation}
This mass squared is actually negative, so the fine-tuning problem is no longer an issue. However, we have recreated a time instability corresponding to the characteristic length scale $\xi^{-1}$ of the system. This behavior is clearly visible in Fig.~\ref{fig:mR2} where we have plotted both the full mass squared $m_R^2$ (solid black) and the bare function $1/3F_{,R}$ (dotted black) for the $\alpha R\log{(R/\mu^2)}$ model.

%
%

\section{Summary and Discussion}
\label{sec:summary}

We have shown in this paper that attempts to find stable static solutions with $\gamma_{\rm PPN} \approx 1$ in metric $f(R)$ models, designed to also account for the accelerating expansion of the Universe, lead to a string of constraints on the model parameters. One must find a configuration for which simultaneously $F \equiv \partial f/\partial R$ and $f/R$ remain small compared to one in the interior of the star, where the strength of the gravitational field is built up, while the Ricci scalar traces the matter distribution: $R\approx \kappa \rho$. (See also Ref.~\cite{Hu:2007nk}). In addition, for the configuration to be stable in time, the effective mass term $m_R^2$ for a perturbation in the Ricci scalar needs to be either positive~\cite{Dolgov:2003px}, or if negative, $|m_R|^{-1}$ must be much larger than the size of the physical system under consideration. Furthermore, we showed that unless $m_R^2 \lesssim 1/r_{\odot}^2$, the domain of GR-like static configurations shrinks to essentially a point in the phase space, while for example a continuum of solutions corresponding to $\gamma_{\rm PPN} \approx 1/2$ exists. 

Hence, in particular for large positive $m_R^2$, it would appear more natural to expect that the metric around a generic star would correspond to $\gamma_{\rm PPN}$ different from one. To make a more definitive statement one should solve the dynamical gravitational collapse of a protostellar dust cloud, which is beyond the scope of this paper however. Moreover, to a degree the problem with the boundary conditions would merely be translated to setting the initial conditions for the collapse. Nevertheless, we believe that we have identified a potential problem in that to make a given metric $f(R)$ model credible, one should give an argument as to why the GR-like configurations should be preferred. Otherwise, if the PPN and stability constraints are supplemented by our fine-tuning argument, it seems unlikely that any $f(R)$ model can pass the test -- unless one gives up the hope that the theory is also responsible for the accelerating expansion of the Universe. This is because the condition $m_R^2 \lesssim 1/r_{\odot}^2$, combined with $m_R^2 \sim 1/3F_{,R}$, implies that $F_{,R}$ needs to be large. However, since at the same time one needs $F \ll 1$, a tension is created that makes finding a suitable function $f(R)$ very difficult.

Let us finally note that while completely GR-like configurations are hard to construct in $f(R)$ models, it does not mean that such theories would be somehow fundamentally ill. In fact many metric $f(R)$ theories could describe gravitational physics quite well in most situations; it is the very precise information on $\gamma_{\rm PPN}$ from Solar System experiments which eventually forces one to set $R \approx \kappa \rho$.  Indeed, if one looks even at the simplest model with $f(R) = -\mu^4/R$, one finds that setting $R$ essentially to any other value than $\kappa \rho$ gives $\gamma_{\rm PPN} \approx 1/2$~\cite{Kainulainen:2007bt}. Moreover, whenever one has $R \sim \mu^2$, one finds $m_R^2 \sim \mu^2$, so that these configurations are effectively free of any stability problems. Thus, one sees that the Dolgov-Kawasaki instability and fine-tuning problems depend not only on the theory, but also on the particular given \emph{configuration} within a given model. Even in the simplest $-\mu^4/R$ model one can construct sufficiently stable spacetimes for stellar objects that are consistent with the accelerating expansion of the Universe; these are only excluded from describing reality by the very precise PPN constraints.

%
%

\begin{acknowledgments}
This work was partly (DS) supported by a grant from the Finnish Cultural Foundation.
\end{acknowledgments}

%
%

\end{document}